# A MapReduce based Big-data Framework for Object Extraction from Mosaic Satellite Images

Süleyman Eken[1] and Ahmet Sayar[1]

[1]Department of Computer Engineering, Kocaeli University, Kocaeli, Turkey

{suleyman.eken, ahmet.sayar}@kocaeli.edu.tr

## 1 RESEARCH PROBLEM

Earth observation satellites survey the earth by taking pictures along their pre-defined orbit, and send those pictures to ground stations. After having completed their rotations around the earth they create a set of pictures defining and depicting earth's surface. The pictures satellites take are called tiles or mosaics. Most of the time mosaics cannot represent objects as a whole, and they need to be stitched to get a picture of a specific object. Stitching of satellite mosaics is used in many science domains such as remote sensing, geomatics, information science, geophysics, map engineering and cartography, and in many application areas such as object recognition/extraction/tracing, spatial analysis, topologic analysis, civil and military simulation applications, augmented reality applications etc. (Dawn *et al*, 2010).

When vision of the various parts of the same object falls in different mosaics, they need to be stitched together in order to obtain that object in one picture. This is inevitable in the applications of object recognition and extraction. As a real world example, to get Island of Cyprus in one picture by using LandSat-8 satellite mosaics, we need four mosaics to be stitched together to form a whole picture of Cyprus. For some other objects, we might need to stitch tens or even hundreds of mosaics. As the number and sizes of the mosaics to be stitched increases, the computation resources needed, such as CPU and memory, increases exponentially. This makes stitching in only one computer harder or even impossible (Sayar *et al*, 2014). In central computations, we not only have resource shortage and poor performance problems but also some other problems stemming from the spatio-temporal characteristics of satellite images. That is, partially overlapping satellite images (mosaics) might be taken at different times (temporal) for the same location. This might cause misleading pixel values for the same object scattered on different mosaics, and as a result failure in image stitching and object extraction (Sayar *et al*, 2013). In addition, it is not possible to determine the boundaries of an object presented as a satellite image. Extraction of lakes and islands might be easy. However, rivers, roads, and some other user defined objects are impossible to extract with full automatic systems (Eken and Sayar, 2015a).

Aforementioned problems can be classified into two groups. These are (1) resource and performance problems in central computation, i.e. single machine, for stitching mosaics to extract an object and (2) defining both boundaries of an object and corresponding mosaics on a satellite image. For the second problem, we propose a heuristic approach based on user interactions with reference maps. Boundaries of an object are roughly defined by users' mouse actions. Fine-grained boundaries are calculated automatically by the program itself. Regarding resource and performance problems, we propose a distributed big data framework based on MapReduce approach to enable scalable and high performance image stitching and object extraction.

## 2 OUTLINE OF OBJECTIVES

With the rapid development of satellite technologies, satellite images have been used in a variety of purposes more effectively. In this work, we aim to stitch satellite tiles into one exact image and extract interested objects from it by means of proposed distributed big data framework based on MapReduce programming paradigm.

Processing satellite images is harder than processing any other images. Due to the satellite images consist of spatio-temporal data, image stitching is even harder with pixel based methods which are sensitive to the intensity changes, introduced for instance by noise, varying illumination. So, the success rates of these methods are low. Moreover, image stitching usually requires remarkable computational capabilities for large scale applications (Mamta *et al*, 2013). In short, remote sensing (RS) image stitching is data and computation intensive task (Lajiao *et al*, 2015). Some works have been done to stitch RS images in parallel computing paradigm. They are presented in more detail in section 3. Spite of these works, handling large scale images and conducting the parallel programming logic are still serious issues. In this manner, we will be solved large scale RS image stitching with vector representations of the raster RS images. It is expected to give better performance in distributed computations by reducing the negative effects of bandwidth problem. The feasibility and high performance test of the proposed framework will be tested on real satellite images of real objects.

It is planned to achieve the following objectives with our framework enabling scalable and high performance image stitching and object extraction:

- Identifying the object boundaries using a heuristic and semi-automated approach based on user interactions with reference maps (Google Earth etc.),
- Realizing range queries on large amounts of spatial data to detect which RS mosaics to be inputted for image stitching process,
- Testing whether there is enough mosaic images for interested objects or not,
- Stitching of vector representations of detected raster mosaic images in distributed computing model,
- Modelling of stitched image in accordance with Open Geospatial Consortium (OGC) standards such as (point, line, polyline, polygons, etc.),
- Vectorial representations of stitched image/object then can be stored and manipulated through object relational spatial databases. In this way, output of our proposed framework can be used by third parties (other researchers) for spatial and topologic queries.
- The efficiency and feasibility of the proposed system will be examined by various scenarios.

## 3 STATE OF THE ART

With the development of satellite technology, the quantity and quality of satellite images have been increased so much. It is impossible to process these data by conventional methods. Although parallel computing and cloud infrastructure (Hadoop, Hive, HBase, Impala, Spark, and etc.) make the processing of such large data possible, such systems are inadequate for spatial and spatio-temporal data. The way to tackle with large spatial data, such data can either be handled as non-spatial data or extra functions may be added into non-spatial systems. Due to such shortcomings, researchers have developed systems to analyze and handle spatial data in large distributed systems architecture (Ablimit *et al*, 2013; Ahmed and Mohamed, 2014; Ahmed and Mohamed, 2013; Ahmed *et al*, 2015). We will be implemented a framework based on Hadoop (Dean, and Ghemawat, 2010) to extract objects from raster satellite images. The outcome of this work is used as input by researchers working on large spatial data. Moreover, MapReduce structure is performed on images unlike text based data.

Some works utilizing MapReduce programming paradigm have been done on raster images in the literature. Winslett *et al* (2009), MapReduce model to solve two important spatial problems involving vector and raster data, respectively: bulk-construction of R-Trees and aerial image quality computation. Imagery data is stored as compressed DOQQ file format and these files are processed by Mapper and Reducer. Golpayegani and Halem (2009) and Lv *et al* (2010) implement some image processing algorithms using MapReduce model. However, they firstly convert images to text format and then binary format before using them as raw image. So, this preprocessing step is time consuming. Ermias (2011) present processing large scale satellite images based on MapReduce and then give a case study on edge detection algorithms such as Sobel, Laplacian, and Canny. According to research we conducted, low-level image processing operations such as edge detection, noise reduction-removal are usually implemented in distributed RS image processing and medium-level operations are carried out at the very least. Our work is concerned the distribution of high-level image processing operations. In this respect, our framework is different from others.

In addition to processing large scale RS images in distributed manner, we have examined the other issue is a stitching of satellite imagery mosaics. The aim of image stitching is to overlay two or more images according to their common intersecting points/areas. So, larger a 2D view or a 3D representation of the scanned scene might be gained. Image stitching techniques are divided into two classes: (ii) feature-based and (ii) area-based. Area-based methods are preferably applied when the images have not many prominent details and the distinctive information. However, feature-based methods don't consider image intensity values and distribution. Also, there are some stitching methods using simultaneously both area-based and feature-based approaches (Zitova and Flusser, 2003). Lajiao *et al* (2015) give a perspective on the current state of image stitching parallelization for large scale applications. Difficulty and problem of parallel image mosaicking at large scale such as scheduling with huge number of dependent tasks, programming with multiple-step procedure, dealing with frequent I/O operation have been handled.

In aforementioned works, researchers have proposed stitching of few RS images with feature-based or area-based approaches instead of stitching of large scale images. We propose a framework stitching of vector representations of large scale raster mosaic images in distributed computing model. In this way, the negative effect of the lack of resources of the central system and scalability problem can be eliminated. We will be implemented distributed computing architecture providing stitching of large scale RS mosaic images and object extraction with commodity machine cluster as scalable.

# 4 METHODOLOGY

Advanced waterfall method is used as a research methodology (Royce, 1970). It has seven different steps: (i) problem definition, (ii) literature research, (iii) identifying research questions based on state of the art, (iv) proposing answers through a framework, (v) experimental validation by implementing and testing with different scenarios and different data size, (vi) analysing the results found, and (vii) conclusion and future improvement. The thesis is in the fourth step right now.

## 4.1 Overview of the Proposed Framework

Stitching of vector representations of the raster images is expected to give better performance in distributed computations by reducing the negative effects of bandwidth problem. Vectorised images are going to be stitched with two alternative approaches. The first approach is based on point pattern matching/point set registration technique. The second approach is similar to the solution of the longest common sub-sequence problem (LCSP). After having done with stitching and extraction, objects are obtained in the standard vector models such as points, lines, line-strings and polygons with their real coordinate values. In GIS standards are defined by OGC. By this way, extracted objects can be stored and serviced by spatial databases such as PostgreSQL's PostGIS and Oracle-Spatial. Figure 1 illustrates the overview of proposed framework. Detailed information about each step will be given following subsections.

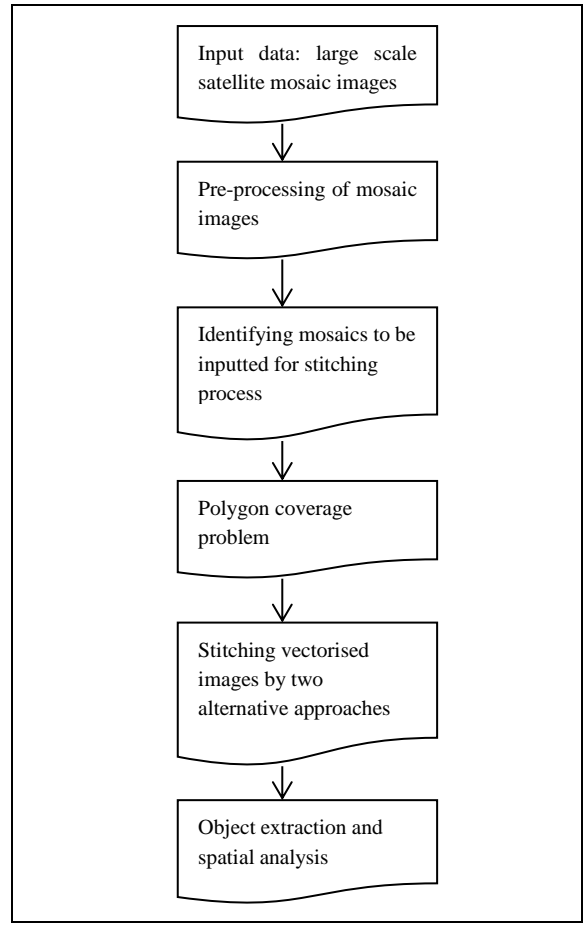

Figure 1: Steps of scalable big data framework

### 4.1.1 Pre-processing of mosaic images

In the first step, high-resolution images to be registered and their metadata such as geographic corner coordinates as NW, NE, SW and SE are obtained from LANDSAT-8 satellite launched by NASA more recently. These mosaic images contain two textures as "USGS" and "NASA". To improve result of the stitching process, these textures are removed from mosaics as pre-processing step.

### 4.1.2 Identifying mosaics to be inputted for stitching process

It is critical problem that which mosaics will be selected for image stitching among big mosaic dataset. In an earlier work, we propose two approaches to overcome mosaic selection problem by means of finding rectangular sub regions intersecting with range query. Former one is based on hybrid of Apache Hadoop and HBase and latter one is based on Apache Lucene. Their effectiveness has been compared in terms of response time under varying number of mosaics. In both approaches, we focused on vertical scalability (different data sizes) instead of horizontal scalability (Eken and Sayar, 2015b).

### 4.1.3 Polygon coverage problem

After identifying mosaics to be inputted for stitching, we must test whether there is enough mosaic images for interested objects or not. It is the process of finding an answer to question of whether a polygon (or spatial query window) is covered fully by given other polygons or not (see Figure 2). Covering problems arise in applications such as telecommunications, spatial query optimization, publish/subscribe middleware, and military sensor coverage and targeting (Daniels and Inkulu, 2001; Giachetta, 2014). For solving polygon coverage problem, we are describing architecture to execute distributed polygon covering algorithm on a Linux cluster using Hadoop MapReduce.

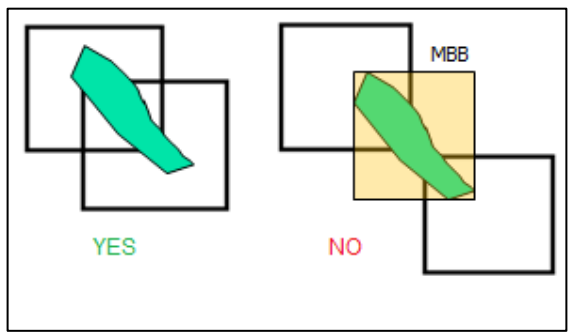

Figure 2: Polygon coverage problem

### 4.1.4 Vectorial based stitching

In this stage, we suppose that we have all mosaics to obtain object specified by user interactions with reference maps. Vector representations of detected raster mosaic images can be now stitched in distributed computing model.

We propose two alternative approaches to stitch large scale satellite mosaic images: (i) point pattern matching/point set registration based and (ii) the second approach is similar to the solution of the longest common sub-sequence problem (LCSP). The second approach is illustrated in Figure 3.

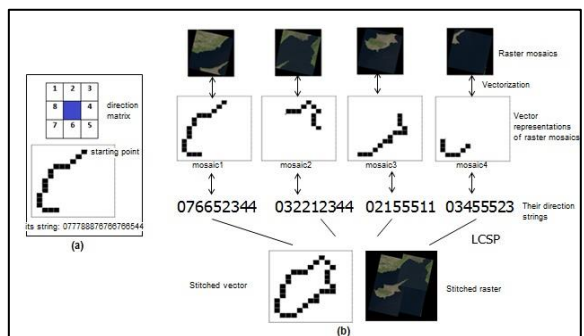

Figure 3: LCSP quasi vectorial based stitching

Apache Hadoop is placed at the core of the framework. To realize such a system, we first need to define mapper and reducer functions, and then their input and output formats. Moreover, in case of having multiple levels of mapper and reducer functions, high level process and data flow needs to be defined. The first level mappers convert raster mosaics into vector counterparts and the following mappers and reducers are performed on the vector mosaics. In near future, we will be sharing information about two approaches in detail.

### 4.1.5 Object extraction and spatial analysis

Spatial databases store and query data for spatial objects in accordance with object relational database model. While typical databases can only store and manipulate various numeric and character types of data, spatial databases can analyse spatial data types such as point, line, and polygon. For the proof of correctness and analysis of the proposed technique, PostgreSQL and its PostGIS extension will be used.

## 5 EXPECTED OUTCOME

The outcomes of this study present valuable knowledge about image processing with distributed scalable big data frameworks. The product obtained by this study can be used in applications requiring spatial and temporal analysis on big satellite map images. This study also shows that big data frameworks are not only used in applications of text-based data mining and machine learning algorithms, but also used in applications of algorithms in image processing. The effectiveness of the product realized with this project is also going to be proven by scalability and performance tests performed on real world LandSat-8 satellite images.

## 6 STAGE OF THE RESEARCH

This paper provides with the background of the research that will investigate into distributed and scalable big data framework for stitching of mosaic satellite images and object extraction. It has explained the motivation of the research and the methodologies and plan of work to be undertaken. The current stage of the research is focusing initially on the first three stages.

The next stage of this research will focus on stitching vectorised mosaic images. Ultimately, as mentioned above, the goal of this stage of the work is to develop a solution to stitching problem. In this way, output of our proposed framework can be used by third parties (other researchers) for spatial and topologic queries.

## Acknowledgments

This work has been supported by the TUBITAK under grant 215E189.